\documentclass[aps,prl,twocolumn,groupedaddress]{revtex4}
\usepackage{graphicx}
\usepackage{dcolumn}
\usepackage{bm}

\begin{document}


\title{Particle-hole symmetry in the antiferromagnetic state of the 
cuprates.\footnote{{\em Expanded version of} Yayu Wang and N. P. Ong, Proc. Nat. Acad. 
Sciences {\rm\bf 98}, 11091 (2001)}}


\author{Yayu Wang, and N. P. Ong}
\affiliation{Department of Physics, Princeton University, Princeton, New Jersey 08544, 
U. S. A.}


\date{\today}


\begin{abstract}
In the layered cuprate perovskites, the occurence of high-temperature superconductivity 
seems deeply related to the unusual nature of the hole excitations.  The limiting case 
of a very small number of holes diffusing in the antiferromagnetic (AF) background may 
provide important insights into this problem.  We have investigated the transport 
properties in a series of crystals of $\rm YBa_2Cu_3O_y$, and found that the temperature 
dependences of the Hall coefficient $R_H$ and thermopower $S$ change abruptly as soon as 
the AF phase boundary is crossed.  In the AF state at low temperatures $T$, both $R_H$ 
and $S$ are unexpectedly suppressed to nearly zero over a broad interval of $T$.  We 
argue that this suppression arises from near-exact symmetry in the particle-hole 
currents.  From the trends in $R_H$ and $S$, we infer that the symmetry is increasingly 
robust as the hole density $x$ becomes very small ($x\simeq 0.01$).  

\end{abstract}
\maketitle                   


\section{Introduction}
In the layered cuprate perovskites, high-temperature superconductivity appears when a 
moderate density of holes are introduced into the $\rm CuO_2$ planes by chemical doping.  
The parent (undoped) compound is a Mott insulator with a spin-$\frac12$ moment residing 
on each Cu ion in the plane.  Below the N\'eel temperature, long-range, 
three-dimensional antiferromagnetism is stable.  The introduction of holes destroys the 
static 3D antiferromagnetic (AF) order, and superconductivity appears when the relative 
hole population $x$ exceeds $\sim$0.05 per Cu ion.  In the quest to understand cuprate 
superconductivity, it has long been recognized that determining the nature of the 
itinerant hole moving in a sea of disordered spin-$\frac12$ moments with short-range AF 
order is an essential step.  Because semi-classical approximations are not reliable for 
this spin-$\frac12$ system, the problem has remained intractable despite a decade of 
theoretical effort \cite{Anderson,Siggia,Wen,Weng,Senthil,Laughlin}.  In the dilute 
limit ($x<0.05$), the hole has been variously described as a `holon' moving in a 
resonating valence bond liquid \cite{Anderson}, or an excitation dressed with a spiral 
spin-current backflow~\cite{Siggia}, or as a charge trailing a `phase string' 
\cite{Weng}.  

In this limit, the signals from the itinerant carriers become increasingly difficult to 
resolve from background signals, especially in experiments that `see' all the valence 
electrons in the sample.  Transport experiments, which probe directly the itinerant 
holes without background problems, would appear to have a valuable role to play.  To 
date, however, the available transport data in the very underdoped limit come almost 
entirely from the single-layer cuprate $\rm La_2Sr_xCuO_4$ (LSCO).  The resistivity and 
Hall results in LSCO do not appear to be qualitatively different from those in the 
moderately underdoped regime ($0.05 < x < 0.13 $) \cite{Takagi1,Hwang}.  However, 
because doping by Sr-substitution creates strong disorder, there is good reason to 
suspect that results in LSCO reflect the properties of the disorder-dominated $\rm 
CuO_2$ plane.

Doping-induced disorder is much less of a factor in the bilayer cuprate $\rm 
YBa_2Cu_3O_y$ (YBCO) because chemical doping proceeds by varying the oxygen content in 
the chains.  In the phase diagram of YBCO, the N\'eel state with 3D long-range 
antiferromagnetism exists in a broad range of oxygen concentration $6.0<y< 6.40$ 
\cite{Tranquada1,Rossat,Andersen}.  Above 6.40, long-range static AF order is suppressed 
by the motion of holes (however, intriguing evidence for a slowly fluctuating 
commensurate AF magnetization has been reported \cite{Keimer} in a crystal with $y$ = 
6.50).  Despite the large range accessible by oxygen doping, the AF state is little 
explored by transport experiments.  Recently, `$d$-wave' anisotropy in the in-plane 
magnetoresistance has been observed (at $y= 6.30$) and interpreted in terms of 
stripes~\cite{Ando1}.  By and large, however, transport results in this regime are very 
sparse.  Previously, the systematic variation of $R_H$ and $\rho$ in YBCO was reported 
by Ito {\em et al.} \cite{Ito} for crystals with $y>$ 6.40.

In extending Hall and thermopower measurements deep into the AF regime in YBCO, we have 
found that the Hall coefficient and thermopower profiles change remarkably as soon as we 
cross the phase boundary at $y$ = 6.40.  Within the AF state, we find that the 
electronic state is characterized by a very robust type of particle-hole symmetry that 
causes both the Hall coefficient $R_H$ and in-plane thermopower $S$ to decrease strongly 
to zero below $T_N$.  Our results on $R_H$ and $S$ imply a self-adjusting mechanism that 
maintains this cancellation over the whole low-temperature region of the AF phase.  The 
existence of particle-hole symmetry places strong constraints on models for the 
electronic state in the AF regime.  

\section{Experimental Results}
YBCO crystals were grown with near-optimal oxygen content in a $\rm Ba_2CuO_x$ flux 
using yttria-stabilized zirconia (YSZ) crucibles.  To tune the oxygen content to a 
particular value $y$, we first attached 3 pairs of electrical contacts on each crystal, 
using EPO-TEK H20E silver epoxy, and then sealed the crystals in a quartz tube with 
crushed polycrystalline $\rm YBa_2Cu_3O_y$ previously prepared with the desired $y$.  
The crystals (of thicknesses 10-30 $\mu$m) were initially annealed at 550 C for a week, 
and then slowly cooled (at 30 C/hr.) to 150 C, and held there for 3 weeks to optimize 
chain ordering \cite{Andersen,Veal}.  The final cooling to 25 C takes $\sim 12$ hr. (we 
did not detwin any of the crystals).  [The low-$T$ annealing at 150 C seems crucial to 
our observations.  A common practice in the past is to work with crystals that have been 
quenched directly from $\sim$520 C to room temperature (quenching has the merit of 
providing a sharp resistive transition in crystals with $y<6.50$).  In the range 
$6.30<y<6.40$, however, the quenched disorder produces a $T_c$ that is strongly 
suppressed, often to below 4 K, and a non-metallic $\rho$ profile.  Subsequent `shelf' 
annealing at room-temperature produces an upward creep of $T_c$ as equilibrium is slowly 
restored (for $y$ = 6.41, for instance, $T_c$ creeps from 13 to 27 K over a 6-day 
period) \cite{Veal}.  It appears that quenching leads to a large degree of in-plane 
disorder which causes localization and strong $T_c$ suppression.  However, the disorder 
can be removed by low-$T$ annealing.  The upward creep in $T_c$ is {\em not} observed in 
our annealed crystals.]  

A recent detailed calibration (based on comparing the thermopower profiles of YBCO and 
LSCO in the underdoped regime) \cite{Wangcalib} finds that the in-plane hole density $x$ 
in YBCO varies linearly with $y$ in the underdoped regime as 
\begin{equation}
x \simeq 0.4(y-y_0), \quad(y< 6.70, y_0\simeq 6.20).
\label{xy}
\end{equation}

\begin{figure}
\includegraphics[width=7cm]{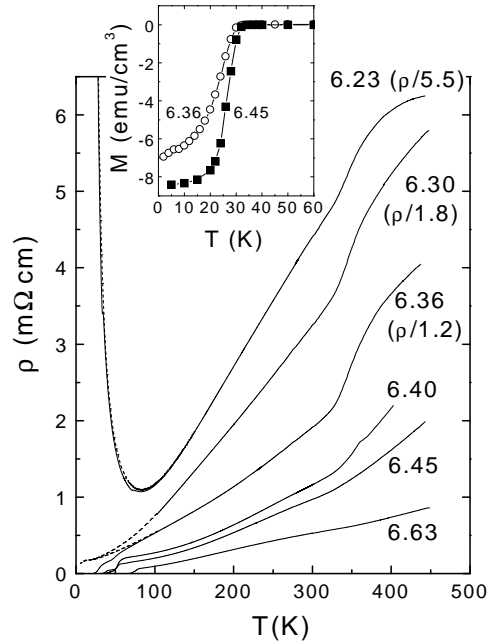}
\caption{The in-plane resisitivity $\rho$ of $\rm YBa_2Cu_3O_y$ crystals measured in 
Samples 1 ($y$ = 6.23), 2 (6.30), 3 (6.36), 4 (6.40), 5 (6.45), 6 (6.63) (the data in 
Samples 1-3 are reduced by the factors shown).  Broken lines for Samples 1-3 represent 
$\rho$ in a 14-T field $\bf H\parallel {\hat c}$ (in Sample 1, the broken line nearly 
coincides with the zero-$H$ data).  The inset shows the magnetization curves taken with 
$H$ = 10 Oe in Samples 3 and 5. 
}
\label{rho}
\end{figure}

In the range $6.23<y<6.63$, we measured in the same crystal the 3 in-plane quantities, 
resistivity $\rho$, Hall coefficient $R_H$ and thermopower $S$ (Samples 1-6). $S$ and 
$\rho$, but not $R_H$, were measured in samples with larger $y$ (Samples 7-9).  [We have 
also added data from a new crystal (Sample 0 with $y$= 6.19) with hole density about 
half that in Sample 1.]  $R_H$ was typically measured with an AC current by slowly 
sweeping the applied field $H$ from -8 T to +8 T and back with $T$ fixed (for all 
samples here, $R_H$ is $H$-independent).  Above 250 K, the swept-field Hall data were 
supplemented by high-density data obtained by a generalized van der Pauw method (fixing 
$H$ at 14 T and alternating the current and Hall voltage contacts (where they overlap, 
the two data sets agree within our uncertainty).  $S$ was measured with a thermal 
gradient of 0.5 K per mm. We corrected for the small contribution from the Au leads.

Figure \ref{rho} displays the in-plane resistivity for several of the crystals measured.
In Sample 1 which has the second smallest $y$ (6.23), $\rho$ increases in a nearly 
activated way as $T$ decreases below 70 K, implying strong localilzation of the 
carriers.  At high temperature (above 320 K), $\rho$ saturates at 38 $\rm m\Omega cm$ 
(corresponding to a mobility $\mu\sim 2\;{\rm cm^2/Vs}$).  Between these two limits, 
$\rho$ varies almost linearly with $T$.  All 3 features qualitatively resemble those 
observed in very underdoped LSCO ($x= 0.04$) \cite{Takagi1} (the magnitudes are also 
comparable if we scale by the hole density given by Eq. \ref{xy}).  Samples 2 and 3 also 
show the $T$-linear variation, but strong localization is not observed at low $T$ (see 
below).

\begin{figure}
\includegraphics[width=7cm]{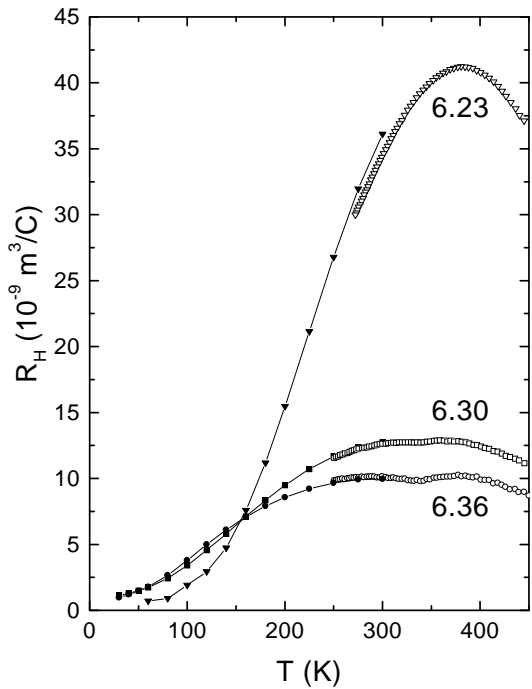}
\caption{The in-plane Hall coefficient $R_H$ of $\rm YBa_2Cu_3O_y$ crystals measured in 
Samples 1-3 with $y$ = 6.23, 6.30 and 6.36, respectively.  In Sample 1, $R_H$ attains a 
broad maximum near the N\'eel temperature $T_N$ and decreases by $\sim$ 50 below 70 K.  
Samples 2 and 3 have a similar behavior except that the low $T$ values are not as 
strongly suppressed.  A slight minimum (`notch') occurs near $T_N$ in 2 and 3, but not 
in 1. In each curve, the high-density data above 250 K were taken using a technique of 
alternating the current and voltage contacts with $H$ fixed at 14 T.  Data with wider 
spacing were taken by slowly scanning the field between -8 and +8 T and back at fixed 
$T$.  
}
\label{LoHall}
\end{figure}

Interestingly, as we cross the AF phase boundary, the $T$-linear dependence abruptly 
changes to the familiar `$S$-shaped' profile that characterizes $\rho$ in the doping 
range $6.40<y<6.70$.  In this range, our $\rho$ are in quantitative agreement with the 
data of Ito {\em et al.}~\cite{Ito}.  

In the narrow range $6.30<y<6.40$ just inside the AF regime, $T_c$ (determined by flux 
expulsion) is highly sensitive to $y$.  As shown in the inset of Fig. \ref{rho}, $T_c$ = 
in Sample 3 ($y$ = 6.36) equals 30 K, compared with 27 K measured by Veal {\em et al.} 
\cite{Veal} in a crystal with $y$ = 6.38, after long-time relaxation.  However, a lower 
$T_c$ (12 K) has been observed by Bonn {\it et al.} in a high-purity, detwinned crystal 
($y$ = 6.35).  Moreover, in Samples 2 and 3, $\rho$ and $S$ show evidence for 
paraconducting fluctuations starting at 70 K.  Because of this variability, we will not 
base any of our conclusions on the behavior of $\rho$ in Samples 2 and 3 in zero $H$ 
below 70 K.  [We note, however, that in a 14-Tesla field $\bf H\parallel {\hat c}$, 
$\rho$ in Samples 2 and 3 saturates to a nearly $T$-independent value as bulk 
superconductivity is suppressed (broken lines in Fig. \ref{rho}).  We do not observe the 
upturn in $\rho$ associated with localization, in sharp contrast with Sample 1.  In 
Sample 1, $\rho$ measured in a 14-T field $\parallel\bf {\hat c}$ (broken line) very 
nearly coincides with the zero-$H$ data.  Provided that the 14-Tesla field is high 
enough to yield the `normal-state' $\rho$ profile in Samples 2 ($x\simeq 0.04$) and 3 
($x\simeq$ 0.064), their metallic behavior implies that the critical value $x_c$ for 
localization in YBCO is much smaller than in LSCO ($x_c\simeq 0.14$).  This conclusion 
is preliminary until we repeat the $\rho$ measurements at much higher fields.]

\begin{figure}
\includegraphics[width=7cm]{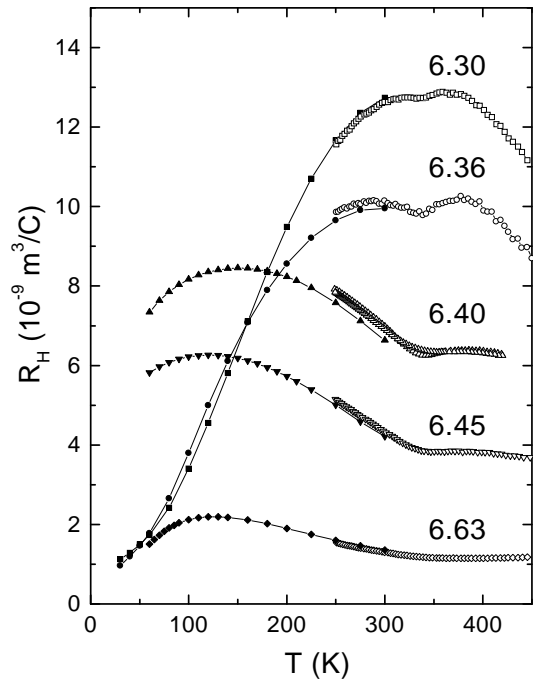}
\caption{The in-plane Hall coefficient $R_H$ of $\rm YBa_2Cu_3O_y$ in Samples 2-6 with 
$y$ = 6.30, 6.36, 6.40, 6.45 and 6.63, respectively.  The profiles of $R_H$ in Samples 2 
and 3 are qualitatively different from those in Samples 4-6.  In Samples 2 and 3, $R_H$ 
peaks near $T_N$ and decreases monotonically as in Fig. \ref{LoHall}.  However, in 
Samples 4-6 ($T_N \sim 0$), the peak temperature in $R_H$ is systematically lower as $y$ 
increases beyond 6.40.  All samples display a `notch' near 350 K where $\rho$ displays a 
slope change. 
}
\label{HiHall}
\end{figure}

\begin{figure}
\includegraphics[width=9cm]{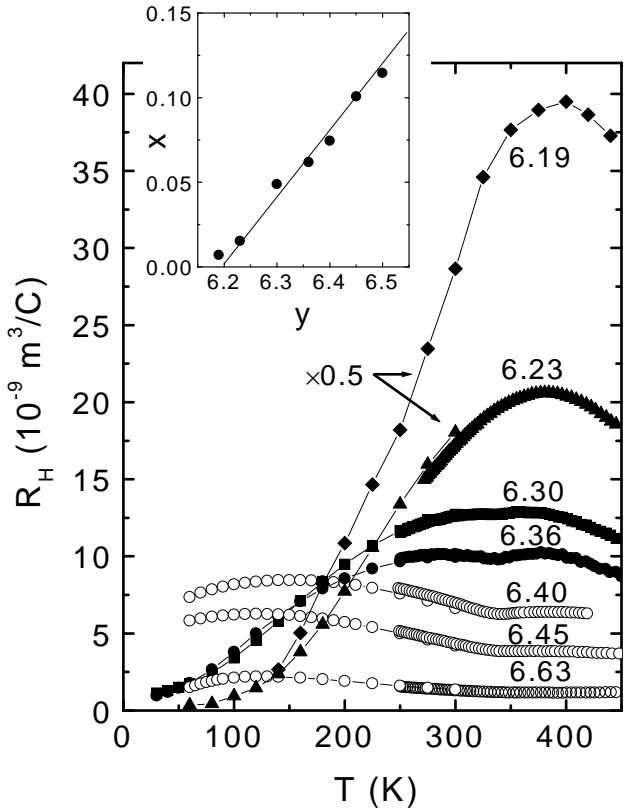}
\caption{Profiles of $R_H$ vs. $T$ in 7 crystals of $\rm YBa_2Cu_3O_y$ Samples 0 ($y$ = 
6.19), 1 (6.23), 2 (6.30), 3 (6.36), 4 (6.40), 5 (6.45), 6 (6.63) (for display purposes, 
the data in Samples 0 and 1 have been reduced by a factor of 2).  Note that between 
Samples 3 and 4, the maximum in $R_H$ shifts abruptly to low $T$.  At the peak, the Hall 
density $n_{H,max} = 1/eR_H$ varies linearly with $(y-y_0)$, as given by Eq. \ref{xy}. 
In the insert, we have plotted as solid circles $n_{H,max}$ (expressed as the equivalent 
$x$) versus $y$. The straight line is Eq. \ref{xy}.  
}
\label{AllHall}
\end{figure}

We describe next the (weak-field) Hall coefficient $R_H$ in these samples.  Figure 
\ref{LoHall} displays the $T$ dependence of $R_H$ in Samples 1-3 (with $y$ = 6.23, 6.30, 
and 6.36, respectively) which lie within the AF region of the phase diagram.  Because of 
the very low carrier densities in these samples, the high-temperature values of $R_H$ 
are dramatically enhanced (10-20 times larger than in optimally-doped YBCO).  At a 
temperature close to the N\'eel temperature $T_N\sim$ 410 K, $R_H$ attains a peak value 
of $\rm 42\times 10^{-9}\; m^3/C$, corresponding to a Hall density $n_H$ of $\rm 
1.50\times 10^{20}\;cm^{-3}$.  The values of $n_H$ at the peak of $R_H$ are in 
quantitative agreement with the calibration in Eq. \ref{xy} as we show later (see Fig. 
\ref{AllHall}).  This implies that, in the lightly-doped regime, the `correct' hole 
density sets the overall magnitude of $R_H$ at temperatures at and above $T_N$.  Below 
$T_N$, $R_H$ decreases steeply, saturating below 70 K to a very small positive value 
nearly $\sim 50$ times weaker than the peak value.  The Hall profiles in Samples 2 and 3 
also show similar decrease below $T_N$.  However, the peak $R_H$ values are smaller 
(consistent with a higher hole concentration) and the low-$T$ values are not as strongly 
suppressed. 

These profiles differ strikingly from the standard $R_H$ profile observed in underdoped 
cuprates at moderately high doping.  In crystals with $y\simeq 6.60$, $R_H$ {\em 
increases} nominally as $1/T$ as the temperature decreases, and attains a maximum at a 
temperature $T_{max,H}\sim$ 110-120 K (this may be the most familiar profile of $R_H$ in 
underdoped cuprates).  

The evolution of the Hall profiles in Samples 1-3 to the pattern seen at higher doping 
is shown in Fig. \ref{LoHall} for Samples 4-6 with $y$ = 6.40, 6.45 and 6.63, 
respectively.

In examining the Hall curves in Fig. \ref{HiHall}, we find that they separate naturally 
into two classes of behavior.  Inside the AF phase (Samples 2 and 3, which are 
reproduced here), $R_H$ reaches a maximum at $T_{max,H}$ close to $T_N$, and then 
decreases below.  The region over which $R_H$ is suppressed extends up to 300 K.  
Samples outside the AF phase (4-7), however, follow a common pattern that is distinct 
from that in the AF phase.  Above the `notch' temperature $T_s\sim$ 350 K (where a 
slight minimum may be observed), $R_H$ is nearly $T$ independent.  [We will discuss the 
notch elsewhere.]  Below $T_s$, $R_H$ increases to a broad peak at $T_{max,H}$ that is 
much lower than in the AF phase.

From Figs. \ref{LoHall} and \ref{HiHall}, we infer the following trend.  Deep in the AF 
state, strong suppression of $R_H$ is apparent over a wide range of $T$ (Samples 1-3).  
However, when the AF boundary is crossed ($y$ = 6.40), this region of suppressed $R_H$ 
and $S$ fills in abruptly.  Concurrently, the peak temperatures in $R_H$ abruptly falls 
from 350 K to 150 K in Sample 4, and to 120 K in Sample 6 ($y$ = 6.63).  

To emphasize this overall trend, we have combined in Fig. \ref{AllHall} the $R_H$ traces 
in Samples 0-6 [we have added data from a new sample (Sample 0, $y$ = 6.19), which has 
the lowest doping level achieved to date].  The insert in Fig. \ref{AllHall} shows that, 
in each crystal, the overall magnitude of $R_H$ as indicated by its maximum value is in 
surprisingly good agreement with the calibration of the hole density given by Eq. 
\ref{xy}) \cite{Wangcalib} for $y<6.60$ (above 6.60 $n_{H,max}$ deviates upwards from 
the straight line).

\begin{figure}
\includegraphics[width=7cm]{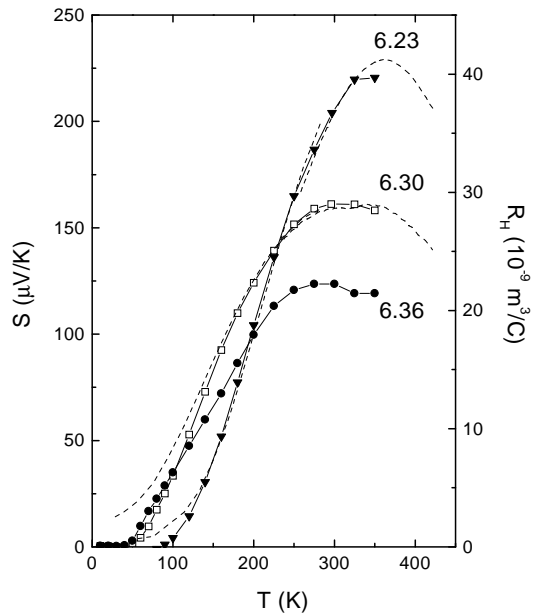}
\caption{The in-plane thermopower $S$ in Samples 1-3 of YBCO (data points).  In Samples 
1 and 2, $S$ decreases monotonically nearly to zero as $T$ decreases below $T_N$ with 
nearly the same $T$ dependence as $R_H$.  The broken lines represent $R_H$ in Samples 1 
and 2 (values of $R_H$ are on the right scale).  In Sample 2 ($y = 6.30$), $R_H$ is 
shown multiplied by 2.25.  
}
\label{Lothermo}
\end{figure}

Remarkably, the pattern of behavior in $R_H$ is matched by the behavior of the in-plane 
thermopower $S$.  Figure \ref{Lothermo} displays $S$ measured in Samples 1-3.  As in 
$R_H$, the thermopower attains a broad peak near $T_N$ and then decreases monotonically 
below $T_N$.  
\begin{figure}
\includegraphics[width=7cm]{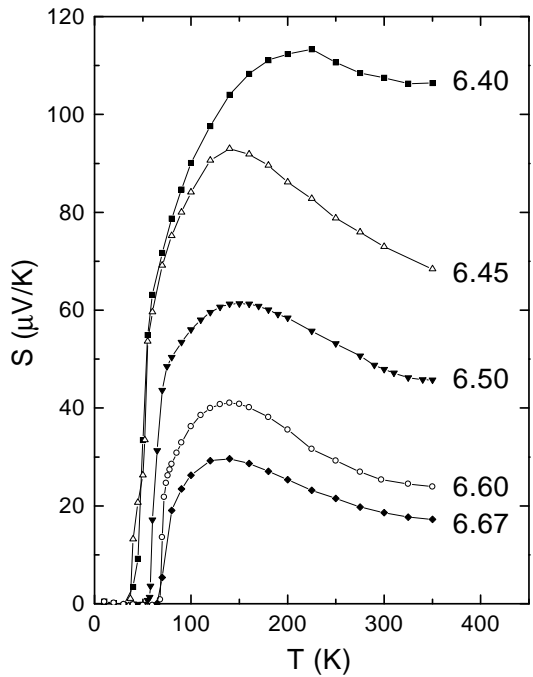}
\caption{The in-plane thermopower $S$ in Samples 4-8 of $\rm YBa_2Cu_3O_y$ (with $y$ = 
6.40, 6.45, 6.50, 6.60, 6.67), which lie outside the AF region.  In each sample, $S$ 
attains a peak at a temperature close to the peak temperature of $R_H$.
}
\label{Hithermo}
\end{figure}
The suppression of $S$ below $T_N$ extends over a broad range of $T$ in these samples.  
A striking feature of $S$ in these samples is that it has virtually the same $T$ 
dependence as $R_H$.  To facilitate comparison, the curves of $R_H$ in Samples 1 and 2 
are replotted as broken lines in Fig. \ref{Lothermo} and scaled to match the 
corresponding curves for $S$.  The comparisons show that both the Hall effect and 
thermopower are suppressed in nearly the same way over a wide range of $T$ inside the AF 
phase.  

We display the thermopower at higher doping in Fig. \ref{Hithermo}.  The gradual 
evolution of the profile of $S$ vs. $T$ in the AF phase to the more familiar profile at 
moderate doping also agrees with the corresponding changes in $R_H$.  For $y <$6.40, the 
peak in $S$ is close to $T_N$, whereas for $y>$6.40, it decreases systematically, 
eventually reaching 120 K at $y $= 6.67.  

The simultaneous suppression of the Hall and thermopower signals is anomalous.  We 
review below the general conditions under which cancellation of these signals is 
observed in conventional metals and semiconductors.

\section{Discussion}

In the simplest case of a single-band isotropic Fermi Surface (FS) with carrier density 
$n$, the weak-field $R_H$ is expected to increase as $1/n$, while $S\sim1/\epsilon_F$ as 
$n$ decreases ($\epsilon_F$ is the Fermi energy).

For our experiment, the important aspect of $R_H$ and $S$ is their sensitivity to the 
sign of the carriers.  When both electrons and holes are present (mixed conduction), the 
observed Hall and thermopower signals are the algebraic sums of contributions from the 
two carrier types. 

In terms of $\sigma$ (conductivity) and $\sigma_H$ (Hall conductivity), $R_H$ equals 
$\sigma_H/\sigma^2H$.  As $\sigma_H$ measures the transverse current produced by the 
Lorentz force in an applied $H$, it is sensitive to the sign of the carrier charge.  For 
mixed conduction, $\sigma_H$ is the sum
\begin{equation}
\sigma_H = \sigma^e_H + \sigma^h_H,
\label{sH}
\end{equation}
where the electron and hole Hall conductivities ($\sigma^e_H$ and $\sigma^h_H$, 
respectively) are of opposite signs.  In a 2D metal with {\em arbitrary} FS shape and 
lifetime anisotropy, $\sigma_H$ scales as $\langle\ell^2\rangle$ (where $\ell$ is the 
mean-free-path and $\langle\cdots\rangle$ denotes averaging over the FS) \cite{2DHall}.   
Exact cancellation between the 2 terms, if any, occurs at discrete accidental 
temperatures where their $\langle\ell^2\rangle$ happen to be equal.  Away from these 
accidental values of $T$, the scale of $R_H$ is of the order $1/ne$ with $n$ set by the 
dominant Fermi Surface (FS) pocket.

While the sign of $\sigma_H$ depends on the FS {\em curvature} of the dominant pocket, 
the thermopower probes particle-hole asymmetry in a rather different sense.  In a 
thermopower measurement, the current generated by an applied temperature gradient 
$-\nabla T$ is cancelled by an opposite current produced by an electric field $\bf E$ 
(produced by charge accumulation at the ends of the sample).  We have
\begin{equation}
\alpha(-\nabla T) = -\sigma {\bf E},
\label{currents}
\end{equation}
where $\alpha$ is the Peltier conductivity and $\sigma$ the electrical conductivity.  
In conventional metals with a single FS, the thermopower $S = -E/|\nabla T| = 
\alpha/\sigma$ is given by 
\begin{equation}
S = \frac{1}{\sigma} \frac{\pi^2 k_B^2T}{3e}
\left(\frac{\partial{\cal G}}{\partial\epsilon}\right)_{\mu},
\label{S}
\end{equation}
where the `conductivity function' ${\cal G}(\epsilon) = e^2D(\epsilon)\langle 
v_x\ell_x\rangle$ is essentially the integrand of $D(\epsilon)$ weighted by the product 
$v_x\ell_x$.  Hence, $S$ may vanish if ${\cal G}$ displays exact particle-hole symmetry 
about $\mu$.  (If we have separate particle and hole FS pockets, $S$ may also vanish by 
accidental cancellation as discussed for $\sigma_H$.)

Unlike the case for $\sigma_H$, full cancellation leading to a zero $S$ is rarely 
observed, especially in the low-density limit.  When $S$ is observed to cross zero, as 
at low $T$ in noble metals (e.g. Au), the cause is an extraneous effect (phonon drag) 
rather than a matching of the derivatives of ${\cal G}(\epsilon)$.  The present results 
are all the more unusual because $R_H$ and $S$ share nearly the same $T$ dependence over 
such a broad range of $T$.  Some of the issues are aired by analyzing the following 
situations. 

It is fair to ask if, in the most underdoped samples, the holes undergo charge 
segregation and are confined to {\em macroscopic} high-conductivity regions that are 
connected along the sample length.  We argue that the local nature of current 
cancellation (Eq. \ref{currents}) in a thermopower measurement rules out this scenario.  
The high-conductivity channel may be modelled as a self-avoiding chain of $N$ segments 
(of conductivities $\alpha_M$ and $\sigma_M$) described by ${\bf d}_j\equiv d_j{\bf 
\hat{e}}_j$.  Applying Eq. \ref{currents} to each segment, we have 
\begin{equation}
{\bf E}_j\cdot{\bf d}_j = -\frac{\alpha_M}{\sigma_M}(-\vec{\nabla} T)\cdot {{\bf d}}_j.
\label{S1d}
\end{equation}
As the sum of ${\bf E}_j\cdot{\bf d}_j$ over $j$ is the observed Seebeck voltage, Eq. 
\ref{S1d} implies that the observed thermopower takes on the value of the 
high-conductivity material, i.e. the conducting channel largely determines the Seebeck 
voltage.  We believe that the sharp differences between the profiles of $S$ in Samples 1 
and 5 (say) precludes this scenario.

A related issue is whether localization plays a role in the suppression of $R_H$.  In 
conventional systems, the onset of Anderson localization either leaves $R_H$ unchanged 
through the metal-insulator transition (Si:P and $\rm In_2O_3$) or to an {\em increase} 
in $R_H$ (Si:B and Ge:Sb \cite{Sarachik}), i.e. the opposite of what is needed to 
account for the present data.  In underdoped LSCO (where the onset of strong 
localization is most heavily studied), localization also leads to an increasing $R_H$ at 
low $T$ \cite{Xu}.  We are not aware of a material in which an increase in $\rho$ 
(driven by localization) is accompanied by a plummeting $R_H$.  Quite apart from these 
general remarks on the behavior of $R_H$ in the presence of localization, we note that 
the regime of localization in YBCO is quite restricted.  In Sample 1 it onsets at 70 K 
whereas $R_H$ and $S$ start declining at 300 K (Fig. \ref{LoHall}), so these can hardly 
be related phenomena.  Similarly, in Samples 2 and 3, the behavior of $\rho$ in an 
intense field (Fig. \ref{rho}) sets a conservative upper bound of $\sim$20 K for the 
onset of localization, whereas $R_H$ and $S$ start falling at 300 K.  We conclude that 
the suppression of $R_H$ is unrelated to localization.

Finally, we note that a quasi-one-dimensional (1D) electronic state in itself does not 
ensure a zero $R_H$ (see also Ref. \cite{Emery}).  Quasi-1D conductors whose Hall effect 
have been measured typically display a finite and strongly $T$ dependent $R_H$ 
\cite{Tessema} because a very small but measurable hopping amplitude transverse to the 
chains is sufficient to produce a finite Hall resistivity (the low carrier densities in 
these systems also strongly amplify $R_H$).    

\section{Particle-Hole Symmetry}
With the above remarks, let us discuss $R_H$ and $S$ in Samples 1-3 (Figs. \ref{LoHall} 
and \ref{Lothermo}).  The overall scale of $R_H$ in Figs. \ref{LoHall} and \ref{HiHall} 
(as indicated by its peak value) increases rapidly as $y$ decreases to 6.20.  
Interestingly, we find that the Hall density $n_H = (eR_H)^{-1}$ evaluated at the peak 
is in very good agreement with the calibration Eq. \ref{xy}.  Hence, $R_H$ starts 
falling from a value fixed by the hole density near $T_N$, and then decreases virtually 
to zero.  The decrease is matched by that in $S$.  In this state, particle-hole symmetry 
of a rather unusual type appears, and becomes increasingly robust with decreasing $T$. 

\begin{figure}
\includegraphics[width=7cm]{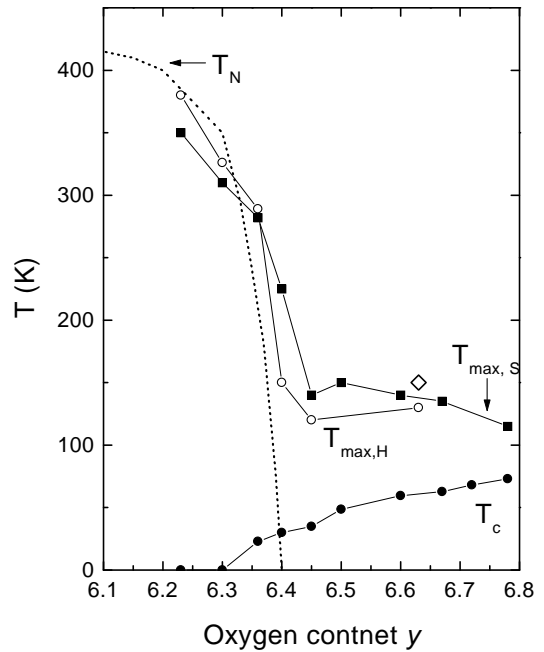}
\caption{Dependence on $y$ of the peak temperatures of $R_H$ ($T_{max,H}$, open circles) 
and $S$ ($T_{max,S}$, solid squares) in $\rm YBa_2Cu_3O_y$.  The $y$ dependence of the 
N\'eel temperature $T_N$ is the broken line \protect\cite{Rossat}.  For $y<$ 6.40 the 
peak temperatures in $R_H$ and $S$ follow closely $T_N$.  At higher doping, however, the 
peak temperatures stay significantly higher than $T_c$ (solid circles).  The open 
diamond is the peak temperature of $(T_1T)^{-1}$ \protect\cite{Takigawa}.
}
\label{phasedia}
\end{figure}

As discussed above, we believe that the decrease in both quantities arises from 
cancellation between particle and hole-like currents.  To achieve the cancellation in 
$R_H$, {\em both} the carrier densities and the average mean-free-paths must be closely 
matched in the two pockets over an extended range of $T$.  Moreover, to maintain the 
same cancellation in $S$, the derivatives of ${\cal G}(\epsilon)$ in the two pockets 
must be equal in magnitude.  Hence, the observed behaviors of $R_H$ and $S$ place very 
stringent conditions on the electronic state.  There appears to be a self-adjusting 
mechanism that automatically maintains the symmetry  of the particle and hole currents 
over a broad range of $T$ below $T_N$.  

In the striped phase that appears in Nd-doped LSCO below the LTT-LTO (low-temperature 
tetragonal to orthorhombic) transition at $T_d\sim$ 74 K \cite{Tranquada2,Kivelson}, 
$R_H$ (as well as $\sigma_H$) are observed to decrease towards zero nearly linearly with 
$T$ below $T_d$ \cite{Eisaki}.  $S$ also decreases below $T_d$, but at a faster rate.  
While the earlier discussion of a vanishing $R_H$ was in terms of quasi-1D behavior, a 
recent analysis proposes particle-hole symmetry as the cause \cite{Emery}.  At the hole 
concentration $x = \frac18$ in Nd-doped LSCO, the conducting chains are 
$\frac14$-filled, so that the charge carriers have exact particle-hole symmetry 
(equivalent to a $\frac12$-filled 1D band for spinless fermions).  This readily accounts 
for the rapid suppression of $S$ as well.  There is possibly a close relationship 
between the LTT phase in Nd-doped LSCO and the particle-hole symmetric state in YBCO.  
However, we note that there is a wide disparity in the hole densities.  In Sample 1, the 
estimated $x\simeq 0.01$ is an order of magnitude smaller than that in the Nd-LSCO 
material.  To account for the vanishing of both $R_H$ and $S$, it seems that the 
$\frac14$-filled chain scenario must extend to this low concentration in YBCO.  At such 
low doping levels, however, the chains have to be $\sim$50 lattice spacings apart.  This 
would appear to be a serious problem unless it can be shown that such widely separated 
$\frac14$-filled chains are energetically favorable.  

A competing model in the AF regime is the nodal excitation at the nodes of the pseudogap 
state \cite{Timusk}.  Where in $\bf k$-space do the doped holes end up in the limit 
$x\rightarrow 0$?  In the {\em insulating} oxychloride cuprate $\rm Ca_2CuO_2Cl_2$ 
angle-resolved photoemission spectroscopy (ARPES) experiments \cite{Ronning} reveal that 
the occupied states (in the lower Hubbard band) nearest in energy to the chemical 
potential are located at the points $(\pm\frac{\pi}{2},\pm\frac{\pi}{2})$.  This implies 
that doped holes should initially occupy the nodes of the pseudogap.  Recently, Takagi 
{et al.} have succeeded in growing crystals of $\rm Ca_{2-x}Na_xCuO_2Cl_2$ with Na 
content $x$ = 0.10.  ARPES experiments confirm that the holes indeed occupy short FS 
`arcs' around these nodes \cite{Takagi2}.  Future ARPES experiments on crystals with 
smaller $x$ should determine if the arcs decrease to point-like nodes.  If, in the limit 
$x\rightarrow 0$, the chemical potential is pinned at the cusp of the Dirac cone in the 
`clean' cuprates, we expect the nodal excitations at the bottom of the cone to dominate 
the transport properties.  The symmetry of the density of states about the chemical 
potential then implies a zero $S$, which may account for our thermopower results.  It 
remains to be seen whether the Hall currents cancel in this limit as well.  

\section{Implications outside AF regime}
While we have mainly discussed the matched variation of $S$ and $R_H$ within the AF 
region (Samples 1-3), we find that the two quantities also share common features at 
higher doping.  At doping levels $6.4<y<6.70$, the peak temperatures $T_{max,H}$ and 
$T_{max,S}$  continue to track each other, as the case inside the AF region (Fig. 
\ref{phasedia}).
Within the AF regime ($y<6.40$), the peak temperatures closely follow the $T_N$ line 
\cite{Rossat}.  As we increment $y$ from 6.36 to 6.40, both peak temperatures fall 
steeply together with $T_N$.  Above 6.40, $T_N$ is nominally zero, but $T_{max,H}$ and 
$T_{max,S}$ remain at fairly elevated temperatures (150 to 110 K).  As the plots show, 
the change at 6.40 is quite abrupt.  Interestingly, the maximum (at 150 K) in the $\rm 
^{63}Cu$ nuclear relaxation rate $(T_1T)^{-1}$ for $y = 6.60$ also falls on this line 
\cite{Takigawa} (this is often identified as the pseudogap temperature $T^*$ in YBCO).  
 
While the strong resemblance between the profiles of $R_H$ and $S$ found within the AF 
regime is less evident outside, the profiles remain surprisingly similar in the vicinity 
of the peak temperatures over a large doping regime ($6.40<y<6.70$).  As this correlated 
behavior can hardly be accidental, it raises the following question.  What is origin of 
the peak in $S$ and $R_H$ (and their pronounced decrease below $T_{max,H}$ and 
$T_{max,S}$) the underdoped phase of YBCO that has long been observed?  By comparing the 
behavior with that in the AF regime, we are intrigued by the possibility that the 
cuprate plane is {\em restoring} the particle-hole symmetric state at these high doping 
levels even though {\em static} long-ranged antiferromagnetism is now absent.  The 
`flow' towards a state with particle-hole symmetry seems to be operative even at 
relatively high doping ($6.50<y<6.70$), but its effects on transport only become 
apparent below 150 K.  At these doping levels, strong AF correlation may survive as a 
very slowly fluctuating order that gradually condenses with decreasing $T$ (as observed 
\cite{Keimer} in a recent neutron scattering experiment in a crystal with $y = 6.50$).  
This spin ordering does not strongly affect transport properties until $T$ falls below 
the line at $120-150$ K.  Below this line, however, we recover the particle-hole 
symmetric state in which both $S$ and $R_H$ approach zero.  This new view, that the 
peaks in $R_H$ and $S$ in the underdoped regime reflect the increased dominance of the 
particle-hole symmetric state, invites a careful re-assessment of previous conclusions 
\cite{Cooper} regarding the peaks.

\vspace{3mm}\noindent
{\bf Acknowledgments} We acknowledge support from the U.S. National Science Foundation 
(MRSEC grant NSF-DMR 98-09483), and a grant from the New Energy and Industrial Tech. 
Develop. Org., Japan (NEDO).  We have benefitted from critical comments and helpful 
suggestions from B. Keimer, S. Kivelson, R.B. Laughlin, P.A. Lee, A. Millis, V. 
Muthukumar, Z. X. Shen, H. Takagi, J. Tranquada and S. Uchida.


\begin{thebibliography}{99}
\bibitem{2DHall}
In 2D, for arbitrary FS anisotropy, the weak-field $\sigma_H$ is given by 
$\sigma_H =  \frac{2e^2}{h} \frac{{\bf B}\cdot \vec{\cal A}}{\phi_0}$, where $\phi_0 = 
h/e$ is the normal flux quantum, and 
$\vec{\cal A} \equiv \frac12 \oint d\vec{\ell}_{\bf k}\times\vec{\ell}_{\bf k}$ is the 
directed area swept out by the mean-free-path $\vec{\ell}_{\bf k}$ as the state $\bf k$ 
is taken around the FS.  [Ong, N.P., (1991) Phys. Rev. B {\bf 43}, 193.]
\end{thebibliography}
\end{document}